\documentclass[11pt,a4paper]{article}
\pdfoutput=1
\usepackage{jheppub}
\usepackage{amsmath,amsthm,amssymb,url}
\title{Tunnelling of Pulsating Strings in Deformed Minkowski Spacetime}

\author[1]{Sergio Giardino}

\author[2]{and Victor O. Rivelles}

\affiliation[1]{Departamento de F\'isica  \& Centro de Matem\'atica e Aplica\c c\~oes \\ Universidade da Beira Interior\\
Rua Marqu\^es D'\'Avila e Bolama 6200-001 Covilh\~a, Portugal}
 
\affiliation[2]{Instituto de F\'{i}sica\\ Universidade de S\~{a}o Paulo\\ Caixa Postal 66318, 05315-970 S\~{a}o Paulo, SP, Brazil}

\emailAdd{p12@ubi.pt}

\emailAdd{rivelles@fma.if.usp.br}

\abstract{Using the WKB approximation we analyse the tunnelling of a pulsating string in deformed Minkowski spacetime.}


\begin{document}

\maketitle
\flushbottom

\section{Introduction}

There is a huge body of evidence that the AdS/CFT correspondence holds true for strings in $AdS_5 \times S^5$ and ${\cal N}=4$ super Yang-Mills theory in 4 dimensions. For instance, the energy of spinning and rotating strings matches the anomalous dimension of operators in the gauge theory in the range where they can be compared. Integrability on both sides of the correspondence also provides further support for the correspondence \cite{Beisert:2010jr}. It is also important to test the correspondence in situations with less supersymmetry where the gauge theories have deformed potentials leading to marginally deformed ${\cal N}=2$ or ${\cal N}=1$ supersymmetric gauge theories \cite{Leigh:1995ep,Mauri:2005pa}. The gravitational dual of such theories have a deformed five-sphere characterized by a real parameter $\gamma$ and the dilaton and some RR and NS-NS fields are also present \cite{Lunin:2005jy}. In this situation both sides of the correspondence also have integrable structures \cite{He:2013hxd}. Spinning and rotating strings have also been considered in such deformed context and they confirm the correspondence whenever they can be compared  \cite{Roiban:2003dw,Berenstein:2004ys,Frolov:2005ty,Frolov:2005dj,Beisert:2005if,Gromov:2009tv,Ahn:2010yv,Arutyunov:2010gu,Ahn:2012hs,He:2013hxd,Fokken:2013mza,Wu:2012dj}.

There is a class of string configurations, pulsating strings, which has not received much attention since its dual operator is not completely understood. They have been analyzed in $AdS_5 \times S^5$  \cite{deVega:1994yz,Gubser:2002tv,Minahan:2002rc,Engquist:2003rn,Khan:2003sm,Arutyunov:2003za,Kruczenski:2004cn,Smedback:1998yn,Panigrahi:2012in, Pradhan:2013sja,Arnaudov:2015dea}, $AdS_4 \times CP^3$  \cite{Chen:2008qq,Dimov:2009rd} and other backgrounds \cite{Dimov:2004xi,Bobev:2004id,Arnaudov:2010by,Arnaudov:2010dk,Banerjee:2014bca,Cardona:2014gqa,Panigrahi:2014sia,Banerjee:2015bia}, and more recently they have been studied in the deformed case as well \cite{Giardino:2011jy}.  Since the string presents a periodic motion its dynamics can be characterized by its oscillation number. It is not one of the string charges but it is quite useful to parametrize its behaviour \cite{Kruczenski:2004cn, Beccaria:2010zn, Giardino:2011jy}. At the quantum level it is an adiabatic invariant so it provides information about the semi-classical regime for higher values of the oscillation number. In \cite{Giardino:2011jy} we analysed pulsating strings in deformed Minkowski spacetime and in deformed $AdS_5 \times S^5$ for small deformation. We have found the classical energy in terms of the oscillation number in the high and low energy limits. For high energy we performed the quantization of the highly excited string states to second order in perturbation theory and found that the oscillation number has to be even. In the low energy case we found a new term proportional to $\gamma$  which is not present in the classical case. 

In order to analyse the classical dynamics of the pulsating string we introduced an effective potential which captures all relevant information about the deformed background. When the string pulsates on the deformed five-sphere its effective potential grows smoothly as one of the angles increase from zero to $\pi/2$ and the oscillation number can be expressed in terms of complete elliptic integrals \cite{Giardino:2011jy}. In the case of deformed Minkowski spacetime the string pulsates along the radial direction and the effective potential starts growing from the origin until it reaches a maximum value of $m^2/(2\gamma)$ at $r^2=1/\gamma$ and then goes back to zero far away from the origin (see Fig.1). It is clear that at low energies or small deformation the string has a periodic motion 
that can be quantized perturbatively as done in \cite{Giardino:2011jy}. However, since the potential has a maximum, it is possible for the string to tunnel through the potential barrier and the computation of the transition rate for such a process is the main goal of this paper. 

Non-perturbative phenomenon like tunnelling may be studied semi-classically using the WKB approximation whenever the amplitude or the phase of the wave function is taken to be slowly changing. The WKB method has been applied in several situation involving strings \cite{Berg:1987uw,Brustein:1992qb,Lee:1992ir,Zhao:2006zw,Monin:2008uj} and here it will be used to analyse the behaviour of a pulsating string in deformed Minkowski spacetime. 

This paper is organized as follows. In Section \ref{2} the pulsating string in deformed ten-dimensional Minkowski spacetime will be briefly described. In Section \ref{3} we will use the WKB technique to calculate the transition rate for the pulsating string to tunnel through the potential. In Section \ref{4} we will analyse the classical stability of the pulsating string and show that for small deformation it is stable.  We then present some conclusions in the last section.

\section{Pulsating Strings in Deformed Minkowski Spacetime} \label{2}

The energy of a semi-classical pulsating string in ten-dimensional Minkowski spacetime and in $AdS_5 \times S^5$ was computed in terms of its oscillation number in \cite{Minahan:2002rc} while for the case of a deformed Minkowski spacetime and deformed $AdS_5 \times S^5$ the energy was found in \cite{Giardino:2011jy}. We will briefly review the case of deformed Minkowski spacetime. Lunin and Maldacena \cite{Lunin:2005jy} found a technique to build deformed supergravity backgrounds which have a $U(1) \times U(1)$ global symmetry required by the deformed gauge theory. When applied to the ten dimensional Minkowski spacetime it gives the deformed background
\begin{align}
ds^2&=\eta_{\mu\nu}dx^\mu dx^\nu+\sum_{i=1}^{3}\left(dr_{i}^{2}+\mathcal{G}r_{i}^{2}d\phi_{i}^{2}\right)+\gamma^{2}r_{1}^{2}r_{2}^{2}r_{3}^{2}\mathcal{G}\left(\sum_{i=1}^{3}d\phi_{i}\right)^2, \label{R10_deformado}\\
\mathcal{G}^{-1}&=1+\gamma^2\left(r_1^2r_2^2+r_1^2r_3^2+r_2^2r_3^2\right), \nonumber
\end{align}
where a four dimensional Minkowski spacetime is left undeformed and the remaining six dimensional space with coordinates $(r_i, \phi_i)$, $i=1,2,3$, has a deformation parameter $\gamma$. When $\gamma$ vanishes we recover the ten dimensional Minkowski spacetime. The dilaton and the $B_2$ field have non-trivial configurations given by 
\begin{align}
	B_{2}&= \gamma 
\mathcal{G}\left(r_{1}^{2}r_{2}^{2}\, d\phi_{1}\wedge
  d\phi_{2}+r_{2}^{2}r_{3}^{2}\, d\phi_{2}\wedge
  d\phi_{3}+r_{1}^{2}r_{3}^{2}\, d\phi_{1}\wedge
  d\phi_{3}\right), \nonumber\\
  e^{2\Phi}&=\mathcal{G}  \nonumber.
\end{align}

The parametrization for a pulsating string used in \cite{Giardino:2011jy} is not convenient to analyse stability issues. Instead we will take a string at the origin of Minkowski spacetime with 
\begin{align}\label{2.2}
	t = \sqrt{2} \kappa \tau, \qquad r_1 = r_2 = r(\tau), \quad r_3 = 0, \quad \phi_1 = \phi_2 = m \sigma, \quad \phi_3=0, 
\end{align}
where $m$ is the string winding number. We then get
\begin{align}
 ds^2&= -dt^2+ 2(dr^2+\mathcal{G}\,r^2 d\phi^2), \qquad\\
\mathcal{G}^{-1}&=1+\gamma^2 r^4.
\end{align}
For this choice there is no coupling of the string to the $B_2$ field. Then the 
Nambu-Goto action becomes
\begin{align}\label{2.6}
S=-m \int d\tau\,r\sqrt{\mathcal{G}\left(1-\dot{r}^2\right)}, 
\end{align}
where we set the string tension equal to one.

We can then find that the radial canonical momentum $\Pi$ and the squared canonical Hamiltonian is given by 
\begin{align}
H^2&=\Pi^2+\frac{m^2r^2}{1+\gamma^2r^4}. 
\end{align} 
We can identify an effective potential 
\begin{align}\label{2.8}
V^2(r)=\frac{m^2 r^2}{1+\gamma^2r^4},
\end{align}
which governs the string dynamics. The potential has a maximum at $r=1/\sqrt{\gamma}$ where its value is $m^2/(2\gamma)$ (see Fig 1) providing a barrier for a pulsating string trapped in the region $r<1/\sqrt{\gamma}$ . For a particle of energy $E^2<m^2/(2\gamma)$ there are two points where its radial velocity vanishes
\begin{align}
	R^2_{1,2} = \frac{m^2}{2 \gamma^2 E^2} \left( 1 \mp \sqrt{1 - \frac{4 \gamma^2 E^4}{m^4} } \right).
\end{align}
This means that the pulsating string can, in principle, tunnel from $r<R_1$, through the classically forbidden region of the potential, and escape to the classically allowed region $r>R_2$.

\begin{figure}[h] \label{Fig 1}
 \centering  
       \includegraphics[height=7cm,scale=1]{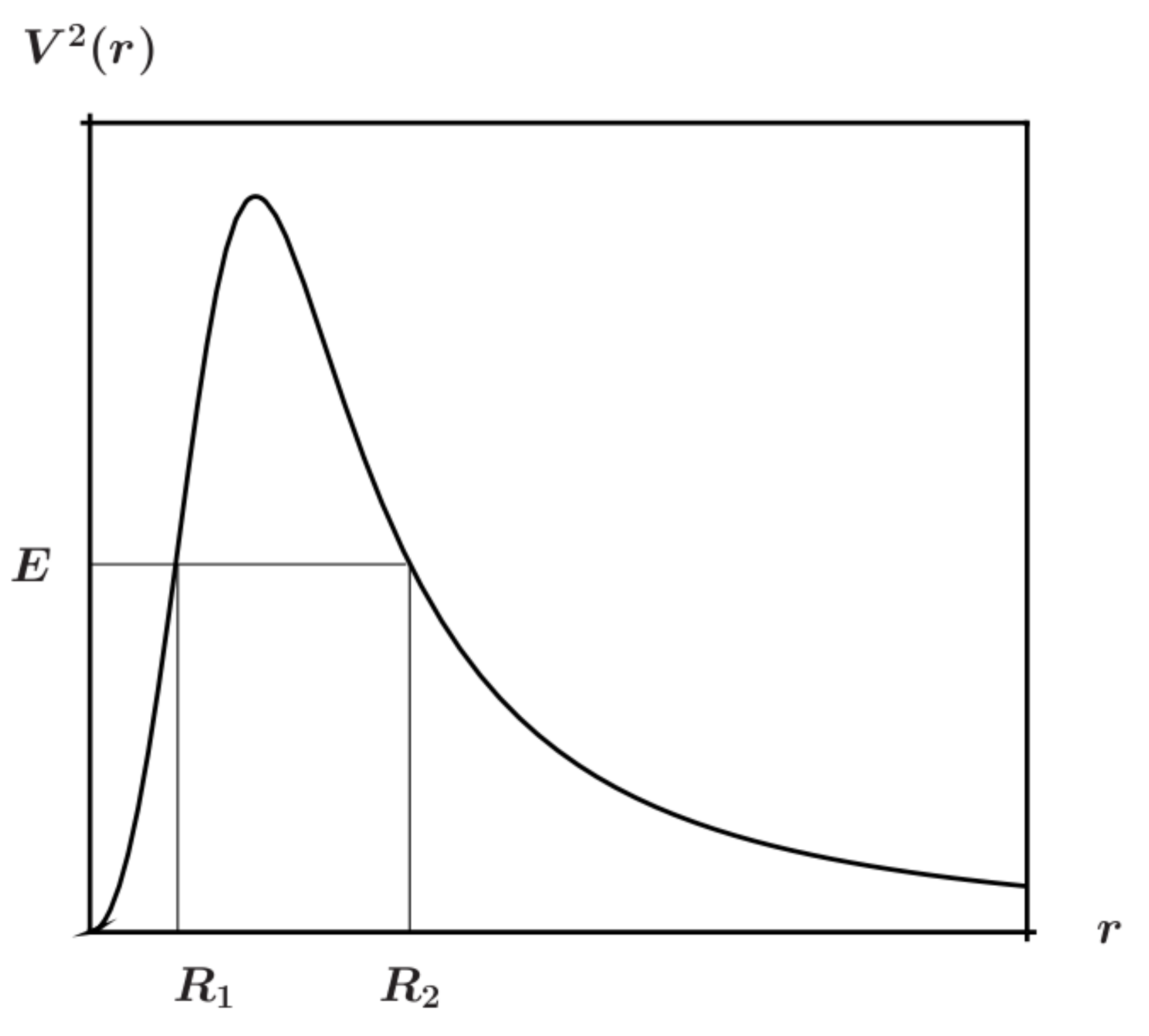}
       \caption{\it The effective potential in deformed Minkowski spacetime}
\end{figure}

The equation of motion can be integrated in terms of elliptic functions and the energy can be found in terms of the oscillation number ${\cal N} = \oint \Pi dr/2\pi$. For more details see \cite{Giardino:2011jy}.

\section{String Tunnelling}\label{3}

To apply the WKB method we assume that the wave function depends only on $r$ so that we can take for $\Pi^2$ the radial part of the Laplacian 
\begin{align}
	\Pi^2 = -\frac{\hbar^2}{\sqrt{-g}}\frac{d}{dr}(\sqrt{-g}\frac{d}{dr}),
\end{align}
where $\sqrt{-g} = r^{(d-1)} \mathcal{G}$ is the determinant of the metric. Here we left the number of dimensions of the deformed part of the space $d$ arbitrary since we want to consider the general situation. For the full ten dimensional case $d=6$. Then the Schr{o}dinger equation reads
\begin{align} \label{3.2}
-\frac{\hbar^2}{r^{(d-1)}}\left(r^{(d-1)}\Psi^\prime\right)^\prime + \frac{4 \hbar^2\gamma^2 r^3}{1 + \gamma^2 r^4} \Psi^\prime + \frac{m^2 r^2}{1+\gamma^2r^4}\Psi= E^2\Psi,
\end{align} 
and the WKB ansatz is 
\begin{align}
\Psi^{WKB}(r) = \left\{
\begin{array}{rl}
\frac{1}{\sqrt{r^{(d-1)} \, \mathcal{G}(r) \, p(r)}} \left( A \, e^{\frac{i}{\hbar}\intop_r^{R_1}dr\, p(r)}+ B \, e^{-\frac{i}{\hbar}\intop_r^{R_1}dr\, p(r)}\right), &  \qquad  r<R_1,\\
& \\
\frac{1}{\sqrt{r^{(d-1)} \, \mathcal{G}(r) \, |p(r)|}}\left( C \, e^{\frac{1}{\hbar}\intop_{R_1}^{r}dr\,|p(r)|}+ D \, e^{-\frac{1}{\hbar}\intop_{R_1}^{r}dr\,|p(r)|}\right),  &  \qquad  R_1 <r < R_2,  \label{WF_WKB}\\
& \\
\frac{1}{\sqrt{r^{(d-1)} \, \mathcal{G}(r) \, p(r)}}\, F \, e^{\frac{i}{\hbar}\intop_{R_2}^{r}dr\,p(r)}, \qquad\qquad\qquad\;\;\;\;\; & \qquad r > R_2,
\end{array}
\right.
\end{align}
where $A, B, C, D$ and $F$ are constants and $p(r) =\sqrt{E^2 - V^2(r)}$. 

The WKB approximation does not hold in the neighbourhood of $r=R_1$ and $r=R_2$ because $p(r)$ vanishes at these points. To avoid this problem we will consider solutions of (\ref{3.2}) around these two points. To this end we introduce coordinates $x = c_i(r - R_i)$, $i=1,2$, where $R_i$ stands for $R_1$ or $R_2$, and we find that (\ref{3.2}) reduces to 
\begin{align}
	\Psi_{R_i}^{\prime\prime} + a_i \Psi_{R_i}^\prime - x \Psi_{R_i} = 0, \label{3.4}
\end{align}
where $\Psi_{R_i}$ is the wave function around $R_i$ and 
\begin{align}
	a_i = \frac{d-5}{c_iR_i} \left( 1 + \frac{4}{d-5} \frac{E^2}{m^2R_i^2} \right). 
\end{align}
If we choose 
\begin{align}
	c_i = \sqrt[3]{\frac{2E^2}{\hbar^2R_i} \frac{1-\gamma^2 R_i^4}{1+\gamma^2 R_i^4}},
\end{align}
we can reduce (\ref{3.4}) to the Airy equation by the change of variable  $\Psi_{R_i}(x) = \exp(-a_i x/2)\chi_i(x)$ which yields 
\begin{align}
\chi_i^{\prime\prime} - \left(x + \frac{a_i^2}{4} \right) \chi_i = 0.
\end{align}
Then, near $R_1$ and $R_2$ we have
\begin{align}
\Psi_{R_i}(x) = e^{-\frac{a_i x}{2}} \left( a_i \, Ai(x+\frac{a_i^2}{4}) + b_i \, Bi(x+\frac{a_i^2}{4})  \right),  
\end{align}
where $a_i$ and $b_i$ are integration constants and $Ai(x)$ and $Bi(x)$ are the two linearly independent Airy functions. 

To match the WKB and the Airy solutions around $R_i$ we must make sure that they have the same functional form for large $|x|$. Around $R_1$ we find that in the WKB solution $\intop_r^{R_1} p(r) dr = 2/3 \hbar (-x)^{3/2}$ for $x<0$, while for $x>0$,  $\intop_r^{R_1} p(r) dr = -2/3 \hbar x^{3/2}$. The Airy functions go like $e^{\pm 2/3 x^{3/2}}$ for $x<0$ and $\cos (2/3 (-x)^{3/2} - \pi/4)$ and $ \sin (2/3 (-x)^{3/2} - \pi/4)$ for $x>0$. Similar expressions hold for the solutions around $R_2$. Matching the solutions we find that around $R_1$ we have 
\begin{align}
	a_1 &= 2 \left(\frac{1+\gamma^2 R_1^4}{\hbar c_1 R_1^{d-1}} \right)^{1/2} \, D, \label{3.9} \\
	b_1 &=   \left(\frac{1+\gamma^2 R_1^4}{\hbar c_1 R_1^{d-1}} \right)^{1/2} \, C, \\
	A &= \left(-i D + \frac{C}{2} \right) e^{i \pi/4},\\
	B &= \left(i D + \frac{C}{2} \right) e^{-i \pi/4},
\end{align}
while around $R_2$ we find
\begin{align}
	a_2 &= 2 \left(\frac{1+\gamma^2 R_1^4}{\hbar c_2 R_2^{d-1}} \right)^{1/2}  e^{ {\cal P}/\hbar} \, C, \\
  b_2 &=   \left(\frac{1+\gamma^2 R_1^4}{\hbar c_2 R_2^{d-1}} \right)^{1/2}  e^{- {\cal P}/\hbar} \, D, \\
	a_2 &= e^{i \pi/4} \left(\frac{1+\gamma^2 R_1^4}{\hbar c_2 R_2^{d-1}} \right)^{1/2} \, F,  \\
	b_2 &= -i a_2, \label{3.16}
\end{align}
where ${\cal P} = \intop_{R_1}^{R_2} |p(r)| dr$. We can now compute $|F/A|^2$ to find
\begin{align}\label{3.17}
	\left| \frac{F}{A} \right|^2 = \left(\frac{1}{4} e^{-{\cal P}/\hbar} - e^{{\cal P}/\hbar} \right)^{-2} = \frac{e^{-2 {\cal P}/\hbar}}{\left( 1 - \frac{1}{4} e^{-2{\cal P}/\hbar} \right)^2}.
\end{align}
Notice that all explicit dependence on $d$ has gone away and only ${\cal P}$ depends on the dimension through the potential $V^2(r)$. 

To find the tunnelling probability we have to consider the probability current $i( \Psi^* \vec{\nabla} \Psi - \Psi \vec{\nabla} \Psi^*)$ in the deformed Minkowski spacetime (\ref{R10_deformado}). Taking only the radial component and integrating it with the proper measure we find that the square root factors in (\ref{WF_WKB}) precisely cancel out the measure factors so that in the region $r<R_1$ it gives $|B|^2 - |A|^2$. Unitarity is then respected since (\ref{3.9}-\ref{3.16}) imply that $|B|^2 - |A|^2 = |F|^2$. This means that the tunnelling amplitude $T=|F/A|^2$ is given by (\ref{3.17}).

The tunnelling amplitude (\ref{3.17}) depends only on ${\cal P} = \int_{R_1}^{R_2} |\sqrt{E^2 - V^2(r)}|  dr$, with $V^2(r)$ given by (\ref{2.8}). This integral is quite complicated but can be performed when the deformation is small. To that end we redefine $r$ as $\tilde{r}= \sqrt{\gamma} r$ so that for $\gamma <<1$ we have
\begin{align}
	\tilde{R}_1 &= \sqrt{\gamma} R_1 = \sqrt{\gamma} \frac{E}{m} \left( 1 + \frac{1}{2} \gamma^2 \frac{E^4}{m^4} \right), \\
	\tilde{R}_2 &= \sqrt{\gamma} R_2 = \frac{m}{\sqrt{\gamma} E} \left( 1 - \frac{1}{2} \gamma^2 \frac{E^4}{m^4} \right),
\end{align}
Notice that $\tilde{R}_1 \tilde{R}_2 = 1$ and from the condition $E^2< m^2/(2\gamma)$ we find that $\tilde{R}_1 <1$. Calling $\tilde{R} = \tilde{R}_1$, we find that 
\begin{align}
	{\cal P} = \frac{E}{\sqrt{\gamma} \tilde{R}} \int_{\tilde{R}}^{1/\tilde{R}} \sqrt{ \frac{(1 - \tilde{R}^2 \tilde{r}^2) (\tilde{r}^2 - \tilde{R}^2)}{1 + \tilde{r}^4}} d\tilde{r}.
\end{align}
We can now split the integral from $\tilde{R}$ to $1/\tilde{R}$ into two integrals, one from $\tilde{R}$ to $1$ and the other from $1$ to $1/\tilde{R}$. For the second integral we can again change the integration variable $\tilde{r}$ to $1/\tilde{r}$ so that 
\begin{align}
	{\cal P} =  \frac{E}{\sqrt{\gamma} \tilde{R}} \int_{\tilde{R}}^1 \left( 1 + \frac{1}{\tilde{r}^2} \right) \sqrt{ \frac{(1 - \tilde{R}^2 \tilde{r}^2) (\tilde{r}^2 - \tilde{R}^2)}{1 + \tilde{r}^4}} \,\,  \tilde{r} \,\, d\tilde{r}.
\end{align}
We can then expand the two factors in the numerator inside the square root and perform the integrals. Keeping only the leading terms in $\gamma$ we find that 
\begin{align}
	{\cal P} = \frac{m}{2\gamma} \left| \ln \left( \frac{\sqrt{\gamma} E}{m} \right) \right|, 
\end{align}
so that when the deformation vanishes the transition amplitude also vanishes as expected.

\section{Classical Stability} \label{4}

As show in the previous Section a pulsating string can tunnel through the potential barrier and this naturally rises questions about its classical stability. It is well known that spinning strings in anti-de Sitter spaces are classically unstable for large spin \cite{Frolov:2003tu}. Pulsating strings, on the other side, have better stability properties than spinning strings as shown in \cite{Khan:2005fc}. In the following we will analyse the stability properties of pulsating strings in deformed Minkowski spacetime. We will apply the technique developed by Larsen and Frolov \cite{Larsen:1993iva} and we will show that when the deformation is small the classical pulsating string is stable. 

We start with the Polyakov action in curved spacetime regarding the string coordinates and the worldsheet metric as independent variables. Following \cite{Larsen:1993iva} the first variation of the Polyakov action gives

\begin{equation}
\delta S_P=-\frac{\sqrt{\lambda}}{2}\int d\xi^2\left[\left(\frac{1}{2}h^{ab}G-G^{ab}\right)\delta h_{ab}-2g_{\mu\nu}\Big(\Box X^\nu+h^{ab}\Gamma^\nu_{\lambda\kappa}X_{,a}^\lambda X_{,b}^\kappa\Big)\delta X^\mu\right],
\end{equation}
where $G_{ab}=g_{\mu\nu}X^\mu_{,a}X^\nu_{,b}$ is the induced metric, $G=h^{ab}G_{ab}$ and $\xi^a$ are the worldsheet coordinates. 
In order to get the second variation of the action, a general perturbation $\delta X^\mu$ is decomposed into normal and tangential components on the worldsheet as 
\begin{equation}
 \delta X^\mu=X^\mu_{,a}\delta X^a+n^\mu_r\delta X^r, \qquad r=2,\dots 9, 
\end{equation}
where $\delta X^a$ is the tangential perturbation and $\delta X^r$ is the normal variation. The normal vectors $n^\mu_a$ are orthonormal to each other and obey
\begin{eqnarray}\label{base}
&&g_{\mu\nu}\,n^\mu_r\,n^\nu_s=\delta_{rs}, \nonumber \\
&&g_{\mu\nu}\, X_{,a}^\mu\, n_r^\nu=0.
\end{eqnarray}
The non-physical perturbations are then excluded by the choice $\delta X^a=0$. We now introduce the second fundamental form $\Omega_{rab}$ and the normal fundamental form $\mu_{rsa}$ defined, respectively, as
\begin{eqnarray}
&&\Omega_{rab}=g_{\mu\nu}\,n^\mu_r\,\nabla_a X^\nu_{,b}, \nonumber \\
&&\mu_{rsa}=g_{\mu\nu}\,n^\mu_r\,\nabla_a n^\nu_s,
\end{eqnarray}
where $\nabla_a=x^\rho_{,a}\nabla_\rho$, with $\nabla_\rho$ being the space-time covariant derivative. After these definitions the second variation of the action is found to be 
\begin{eqnarray}
\delta^2 S_P&=&-\frac{\sqrt{\lambda}}{2}\int d\xi^2\sqrt{-h}\left[\delta h_{ab}\left(2G^{bc}h^{ad}-\frac{1}{2}h^{ad}h^{bc}\,G-\frac{1}{2}h^{ab}\,G^{cd}\right)\delta h_{cd} \right.\nonumber\\
&+&\left. 2\,\delta h_{ab}\,h^{ac}\,h^{bd}\Omega_{rcd}\,\delta x^r 
-2\delta X^r\big(\delta_{rs}\Box-h^{ab}g_{\mu\nu}\nabla_a n^\mu_r\,\nabla_b n^\nu_s-2h^{ab}\mu_{rsa}\,\partial_b \right.\nonumber \\
&-&\left.h^{ab}X^\mu_{,a}X^\nu_{,b}R_{\mu\kappa\lambda\nu}\,n^\kappa_r\, n^\lambda_s\big)\delta X^s\right],
\end{eqnarray}
where $R_{\mu\kappa\lambda\nu}$ is the Riemann tensor. Taking into account that the variation of the internal metric is related to the variation of the spacetime coordinates by
\begin{equation}
 \Omega_{abr}\delta X^r=-\frac{1}{4}\big(G\delta h_{ab}-h_{ab}G_{cd}\delta h_{cd}\big),
\end{equation}
it can be shown that the second variation of the action is 
\begin{eqnarray}
&&\delta^2 S_P=-\frac{\lambda}{2}\int d\xi^2\sqrt{-h}\times\\ \nonumber
&&\delta X^r \left( \delta_{rs} \Box -2h^{ab}\mu_{rsb}\partial_a-\delta^{tu}h^{ab}\mu_{rta}\,\mu_{sub}-\frac{2}{G}h^{ac}h^{bd}\Omega_{abr}\Omega_{cds}-h^{ab}X^\mu_{,a}X^\nu_{,b}R_{\mu\kappa\lambda\nu}\,n^\kappa_r\, n^\lambda_s\right)\delta X^s.
\end{eqnarray}
The equation of motion for the perturbation is then given by 
\begin{eqnarray}\label{motion}
&&\Box\delta X_r-2h^{ab}\mu_{rsa}\partial_b\delta X^s-h^{ab}\nabla_a\mu_{rsb}\delta X^s-\nonumber\\
&&-\left(\delta^{tu}h^{ab}\mu_{rta}\,\mu_{sub}+\frac{2}{G}h^{ac}h^{bd}\Omega_{abr}\Omega_{cds}+h^{ab}X^\mu_{,a}X^\nu_{,b}R_{\mu\kappa\lambda\nu}\,n^\kappa_r\, n^\lambda_s\right)\delta X^s=0.
\end{eqnarray}

Now we will particularize the stability analysis to the deformed Minkowski space (\ref{R10_deformado}) using (\ref{2.2}). 
The equation of motion for $r(\tau)$ is 
\begin{equation}\label{erre}
\dot{r}^2=\kappa^2-m^2r^2\mathcal{G}.
\end{equation}
For small deformation, $\gamma^2\ll 1$, it reduces to $\dot{r}^2 = k^2 - m^2 r^2 + \gamma^2 m^2 r^6$ and it oscillates between $r=0$ and $r=r_-$, where $r_-^2= k^2/m^2 + \gamma^2 k^6/m^6$. Then the motion is periodic with amplitude $\frac{k}{m}( 1 + \frac{1}{2} \gamma^2 \frac{k^4}{m^4})$. We can find explicit solutions like
\begin{eqnarray}\label{sol1}
&&r(\tau) = \frac{k}{m} \left( 1+ \frac{1}{2} \gamma^2 \frac{k^4}{m^4} \right) \sin \left( m\tau + \gamma^2 \frac{k^4}{m^4} g(m\tau) \right), \nonumber \\
&&g(m\tau) = -\frac{15}{16} m \tau + \frac{1}{4} \sin (2m\tau) - \frac{1}{64} \sin (4m\tau),
\end{eqnarray}
but they will not be needed for the stability analysis. 

The induced metric is given by $G_{00}=-G_{11}=-2m^2r^2\mathcal{G}$ and the orthogonality of the normal basis (\ref{base}) requires  
\begin{eqnarray}\label{constraint_def}
&&\sqrt{2}\,\kappa\, n_r^t - ( n^{r_1}_r+n^{r_2}_r) \dot{r} = 0, \nonumber \\
 && n^{\phi_1}_r+n^{\phi_2}_r=0.
\end{eqnarray}
The choice of the normal vectors satisfying the first constraint in (\ref{constraint_def}) demands some work. So let us denote our basis vectors as 
\begin{equation}
n_r^\mu=\left(n_r^t,\,n_r^{x_i},\,n_r^{r_i},\,n_r^{\phi_i}\right), \qquad i=1,\,2,\,3.
\end{equation}
Consider the first constraint for the normal vectors of the form 
\begin{equation}
n_r^\mu=\left(n_r^t,0,\,0,\,0,\,n_r^{r_1},\,n_r^{r_2},\,0,\,0,\,0,\,0\right),
\end{equation}
which are choosen to be non trivial for $r=2,\,3$. Using the constraint and the orthogonality condition, we obtain
\begin{equation}\label{NP}
\left[(1-\alpha^2)n^{r_1}_3-\alpha^2n^{r_2}_3\right]n^{r_1}_2+\left[(1-\alpha^2)n^{r_2}_3-\alpha^2n^{r_1}_3\right]n^{r_2}_2=0,
\end{equation}
where $\alpha=\frac{\dot r}{\sqrt{2}\,\kappa}\leq 1$. The constraint also allow us to rewrite the normalization as
\begin{equation}\label{norm_n}
\left(1-\alpha^2\right)\left(n^{r_1}_r+n^{r_2}_r\right)^2-2\,n^{r_1}_r\,n^{r_2}_r=1.
\end{equation}
Matching the norm of the vectors and using (\ref{NP}), we obtain
\begin{equation}
n_2^{r_1}=\pm\frac{\left(1-\alpha^2\right)n_3^{r_2}-\alpha^2n_3^{r_1}}{\sqrt{1-2\alpha^2}}.
\end{equation}
By setting $n^{r_1}_3=-n^{r_2}_3=\frac{1}{\sqrt{2}}$ 
we finally find 
\begin{eqnarray}
&& n_2=\frac{\kappa}{m\,r\,\mathcal{G}}\left(\frac{\dot r}{\kappa},\,0,\,0,\,0,\,\frac{1}{\sqrt{2}},\,\frac{1}{\sqrt{2}},\,0,\,0,\,0,\,0\right), \nonumber\\
&& n_3=(0,\,0,\,0,\,0,\,\frac{1}{\sqrt{2}},\,-\frac{1}{\sqrt{2}},\,0,\,0,\,0,\,0),\label{base_n}\nonumber\\
&& n_4=(0,\,1,\,0,\,0,\,0,\,0,\,0,\,0,\,0,\,0),\nonumber\\
&& n_5=(0,\,0,\,1,\,0,\,0,\,0,\,0,\,0,\,0,\,0),\\
&& n_6=(0,\,0,\,0,\,1,\,0,\,0,\,0,\,0,\,0,\,0),\nonumber\\
&& n_7=(0,\,0,\,0,\,0,\,0,\,0,\,1,\,0,\,0,\,0),\nonumber\\
&& n_8=(0,\,0,\,0,\,0,\,0,\,0,\,0,\,\frac{1}{\sqrt{(r_1^2+r_2^2)\mathcal{G}}},\,-\frac{1}{\sqrt{(r_1^2+r_2^2)\mathcal{G}}},\,0),\nonumber\\
&& n_9=(0,\,0,\,0,\,0,\,0,\,0,\,0,\,0,\,0,\,\frac{1}{r_3}).\nonumber
\end{eqnarray}
We notice that there are only seven basis vectors since we gauge fixed three coordinates $t$, $\phi_1$ and $\phi_2$. 

The fundamental forms can now be found.  Using the ansatz (\ref{2.2}) we get 
\begin{eqnarray}
&&\mu_{rs0}=0, \nonumber \\ 
&& \mu_{rs1}=mr\mathcal{G}\big[n^{\phi_1}_r(n^{r_1}_s-n^{r_2}_s)-
n^{\phi_1}_s(n^{r_1}_r-n^{r_2}_r)\big], \nonumber\\
&&\Omega_{r01}=0, \nonumber \\
&& \Omega_{r11}=-m^2(n^{r_1}_r+n^{r_2}_r)r\mathcal{G}^2(1-\gamma^2r^4), \nonumber \\
&&\Omega_{r00}=3(n^{r_1}_r+n^{r_2}_r)\ddot r.
\end{eqnarray}
Using the expressions for the basis vectors (\ref{base_n}) and that $\ddot r=m^2r\mathcal{G}(1-2\mathcal{G})$ we obtain the non vanishing components 
\begin{equation}
\mu_{381}=-m\sqrt{\mathcal{G\,}},\qquad\Omega_{200}=3\,\Omega_{211},\qquad\Omega_{211}=\kappa m\sqrt{2\mathcal{G\,}}(1-2\mathcal{G}).
\end{equation}

In order to calculate the curvature term in the equations of motion for the perturbations (\ref{motion}) we use the ansatz (\ref{2.2}) to find the non-vanishing components of the curvature tensor
\begin{eqnarray}
&&R_{r_1\phi_1 r_1\phi_1}=R_{r_1\phi_1 r_2\phi_1}=R_{r_1\phi_2 r_2\phi_2}=R_{r_2\phi_2  r_2\phi_2}=3\gamma^2 \,r^4\mathcal{G}^3, \nonumber\\
&&R_{r_1\phi_2 r_1\phi_2}=R_{r_2\phi_1 r_2\phi_1}= \gamma^2 r^4\big(1-2r^4\gamma^2\big)\mathcal{G}^3,\\
&& R_{r_3\phi_1 r_3\phi_1}=R_{r_3\phi_2 r_3\phi_2}=-R_{r_3\phi_1 r_3\phi_2}= \gamma^2 r^4\mathcal{G}^2,\nonumber\\
&& R_{\phi_1\phi_2 \phi_1\phi_2}=2 \gamma^2 \,r^6\mathcal{G}^4.\nonumber
\end{eqnarray}
The curvature dependent terms then become 
\begin{equation}
G^{ab}X^\mu_{,a}X^\nu_{,b}R_{\mu\rho\sigma\nu}n_r^\rho n^\sigma_s\delta X^s=
\left\{
\begin{array}{ll}
\frac{\gamma^2 \kappa^2}{m^2}(1-6\mathcal{G})\delta X_2,  &r=2\\
- \gamma^2 r^2\mathcal{G}(1-6\mathcal{G}) \delta X_3, &r=3\\
\frac{\gamma^2}{m^2}\big[\dot r^2(8-\mathcal{G})-2\kappa^2\mathcal{G}\big]\delta X_8, &r=8\\
0, &r=4,\,5,\,6,\,7,\,9.
\end{array}
\right.
\end{equation}

Finally, we have to take into account the Kalb-Ramond term
\begin{equation}
S_B= -\sqrt{\lambda}\int d\xi^2 B_{\mu\nu}\epsilon^{ab}X^\mu_{,a}X^\nu_{,b}.
\end{equation}
Its second variation gives
\begin{equation}\label{KRpert}
\delta^2\Big(B_{\mu\nu}\,\epsilon^{ab}X^\mu_{,a}X^\nu_{,b}\Big)=2\gamma \,mr^2G^3\, (3-5r^4\gamma^2)\dot r\Big(\delta X^{r_1}+\delta X^{r_2}\Big)\Big(\delta X^{\phi_2}-\delta X^{\phi_1}\Big).
\end{equation}
Using
\begin{equation}
\Big(\delta X^{r_1}+\delta X^{r_2}\Big)\Big(\delta X^{\phi_2}-\delta X^{\phi_1}\Big)=\Big[\Big(n_2^{r_1}+n_2^{r_2}\Big)\delta X_2+\Big(n_3^{r_1}+n_3^{r_2}\Big)\delta X_3\Big]\left(n_8^{\phi_1}-n_8^{\phi_2}\right)\delta X_8,
\end{equation}
we find the final form for the equations of motion for the perturbations 
\begin{eqnarray}
&&\Big(\Box-\frac{1}{2\,r^2} + \gamma^2 r^2\mathcal{G}(1-6\mathcal{G}) \Big)\,\delta X_3=0,\label{pert_3}\\ 
&&\left(\Box-\frac{5\kappa^2}{4m^2}\frac{(1-2\mathcal{G})^2}{r^4\mathcal{G}}-\gamma^2 \frac{\kappa^2}{m^2}(1-6\,\mathcal{G})\right)\delta X_2 + 4 \gamma \kappa\,\dot r\, \mathcal{G}\,(8\mathcal{G}-5)\,\delta X_8=0,\label{pert_2}\label{EQ1}\\
&&\left(\Box-\frac{1}{2\,r^2}-\frac{\gamma^2}{m^2}\big[(8-\mathcal{G}) \dot{r}^2 -2\kappa^2\mathcal{G}\big]\right)\,\delta X_8 + 4 \gamma \kappa \,\dot r\, \mathcal{G}\,(8\mathcal{G}-5)\,\delta X_2=0, \label{pert_8}\\
&&\,\,\,\,\,\Box\,\delta X_r=0, \qquad r=4,\,5,\,6,\,7,\,9.\label{pert_1}
\end{eqnarray}
Eq.(\ref{pert_1}) shows that for $r=4,\,5,\,6,\,7,$ and $9$ the perturbations are stable so that we have to consider only $r=2,\,3$ and $8$.

From now on we will analyse the stability for small deformation $\gamma<<1$. Keeping only the leading terms in $\gamma$ the equations for the perturbations reduce to 
\begin{eqnarray}
&&\Big(\Box-\frac{1}{2\,r^2}-5 \gamma^2r^2\Big)\,\delta X_3=0, \label{apr_3}\\ 
&&\left(\Box-\frac{5\kappa^2}{4m^2}\frac{1}{r^4}\right)\delta X_2+12 \gamma \kappa\,\dot r\,\delta X_8=0, \label{apr_2}\\
&& \left(\Box-\frac{1}{2\,r^2}\right)\,\delta X_8+12 \gamma \kappa \,\dot r\,\delta X_2=0. \label{apr_8}
\end{eqnarray}
We can now expand $\delta X_i$ as
\begin{equation}
\delta X_i=\sum_{n=-\infty}^\infty e^{in\sigma}F_i(\tau), 
\end{equation}
and use $\Box=\frac{1}{2mr^2}(-\partial^2_\tau+\partial^2_\sigma)$ to get 
\begin{eqnarray}
&&\Big(\frac{d^2}{d\tau^2} + n^2+m^2+10 \gamma^2 m^2r^4 \Big)\,F_3=0, \\ 
&&\left(\frac{d^2}{d\tau^2} +\frac{5\kappa^2}{2}\frac{1}{r^2}\right)F_2 - 24 \gamma \kappa m^2\,r^2\,\dot r F_8=0, \\
&&\left(\frac{d^2}{d\tau^2} +n^2+m^2\right)\,F_8 - 24 \gamma \kappa m^2 \,r^2\,\dot r\,F_2=0. 
\end{eqnarray}
In these equations $r$ is the unperturbed solution to (\ref{erre}) which is a periodic function of $\tau$. Then, by the Sturm theorem, $F_3$ oscillates for large $\tau$ so that the perturbation $\delta X_3$ is stable. We can handle $F_2$ and $F_8$ by expanding then in $\gamma$ as $F_2=U_2+\gamma V_2$ and $F_8=U_8+\gamma V_8$ to get 
\begin{eqnarray}
&&\left(\frac{d^2}{d\tau^2} +\frac{5\kappa^2}{2}\frac{1}{r^2}\right)U_2 = 0, \\
&&\left(\frac{d^2}{d\tau^2} +n^2+m^2\right)\,U_8 = 0, \\
&&\left(\frac{d^2}{d\tau^2} +\frac{5\kappa^2}{2}\frac{1}{r^2}\right)V_2 - 24 \kappa m^2\,r^2\,\dot r U_8=0, \label{4.39}\\
&&\left(\frac{d^2}{d\tau^2} +n^2+m^2\right)\,V_8 - 24 \kappa m^2 \,r^2\,\dot r\,U_2=0. \label{4.40}
\end{eqnarray}
Then $U_2$ and $U_8$ are oscillatory for large $\tau$. In (\ref{4.39}) the homogeneous solution for $V_2$ is also oscillatory for large $\tau$ as is the non-homogeneous term unless $U_8$ has some resonance frequency. This will happens whenever $m^2 = \frac{5k^2}{2r^2}$ or $r^2=\frac{5k^2}{2m^2}$, but because $\frac{5k^2}{2m^2}>r_-^2$ this values of $r$ can not be reached so that there is no resonance. The same  result holds for $V_8$ so the perturbations $\delta X_2$ and $\delta X_8$ are also stable. For arbitrary values of $\gamma$ (\ref{pert_3}) still shows that $\delta X_3$ is oscillatory but (\ref{EQ1}) and (\ref{pert_8}) could not be decoupled.

\section{Conclusions}

We have applied the WKB method to compute the tunnelling amplitude for an oscillating string in deformed Minkowski spacetime. As expected it is proportional to the string energy and vanishes when the deformation goes to zero. We have also shown that for small deformation the classical pulsating string is stable. It is known that pulsating strings in $AdS_5\times S^5$ are dual to operators composed of non-holomorphic products of scalar fields \cite{Engquist:2003rn,Kruczenski:2004cn,Beisert:2003xu,Minahan:2004ds}, but the theory corresponding to the deformed Minkowski spacetime is not known. Since the string tunnelling represents an instability of the system it would be very interesting to find out what happens on the other side of the correspondence.

\acknowledgments
Sergio Giardino is supported by CNPq grant 206383/2014-2 and thanks Prof. Paulo Vargas Moniz and the Center for Mathematics and Applications of the Beira Interior University for hospitality. The work of Victor Rivelles is supported by FAPESP grant 2014/18634-9.

\end{document}